\documentclass[12pt]{article}

\begin{document}

\begin{titlepage}
\title{\vskip .5in
Entropy of Thermally Excited Black Rings}
\author{Finn Larsen\footnote{\texttt{larsenf@umich.edu}} \\
\centerline{\it\normalsize Michigan Center for Theoretical Physics,
Randall Laboratory of Physics}\\
\centerline{\it\normalsize The University of Michigan, Ann Arbor, MI 48109-1120}}

\date{}

\maketitle

\begin{abstract}
A string theory description of near extremal black rings is proposed. 
The entropy is computed and the thermodynamic properties are derived 
for a large family of black rings that have not yet been constructed in 
supergravity. It is also argued that the most general black ring
in $N=8$ supergravity has $21$ parameters up to duality. 
\end{abstract}

\end{titlepage}

\section{Introduction}
\label{sec:intro}
The existence of black ring solutions \cite{Emparan:2001wn} raises 
interesting questions in classical general relativity \cite{Horowitz:2004je} 
and, since black rings are readily embedded 
in supergravity \cite{Elvang:2004rt,Bena:2004de,Elvang:2004ds}, in string theory. 
One important challenge is to understand black rings microscopically in
string theory. For BPS black rings it was proposed that one can 
simply identify circular black rings with straight black 
strings \cite{Emparan:2004wy,Bena:2004tk,Cyrier:2004hj}. This prescription
gives a statistical account of the 
black ring entropy; but it also highlights some confusion about the 
precise distinction between black holes and black rings. Therefore, a 
more detailed microscopic description of black rings is essential even for 
the understanding of black holes. 

The purpose of this paper is to propose a microscopic description 
of thermal black rings, {\it i.e.} rings that are excited 
away from the extremal limit. The strategy is, again, to identify circular black 
rings with straight black strings. It is well-known how to take thermal 
excitations of black strings into account, in the limit where the excitation 
energy is not too large \cite{Callan:1996dv} and we simply adapt this 
description to black rings. The only modest complication is due to the fact 
that black rings carry, in addition to the dipole charges along the circular
string, several additional charges. These charged excitations arise 
as zero-modes of affine currents in the two dimensional effective CFT describing 
the collective excitations of the black string \cite{Larsen:1999uk,Larsen:1999dh}. 
Combining this with general principles, we find an expression for the entropy as 
function of energy, angular momenta, and all charges. Our final result for the
entropy is given in (\ref{micent}) below. As a simple application of this result we
work out the thermodynamic properties of black rings.
 
Thermally excited black rings are much richer than their extremal counterparts. 
This will further focus attention on the limitations of describing black rings in terms 
of black strings. As is clear already in the extremal limit, the identification of these 
two theories is valid only near the objects, where the extrinsic curvature of the black 
ring can be neglected. Such a description is incomplete because, from the dual 
open string perspective, it applies only in 
the infra-red, {\it i.e.} it is an effective description. In particular, this means
the parameters of the microscopic description cannot immediately be identified 
with the corresponding quantities in the supergravity solution (some relevant 
discussions are \cite{Bena:2005ni,Marolf:2000cb}). This is familiar
already for BPS rings where several different definitions of charges and angular 
momenta seem relevant for the description. In the excited theory these ambiguities
persist and, in addition, the microscopic energy must be distinguished from 
the mass of the black ring. Of course, such features do not invalidate the proposed 
microscopic descriptions; they are properties of effective theories.

A by-product of the present investigation is the determination of duality orbits for
black rings. For black holes in five dimensions it is sufficient to consider 
supergravity solutions with $3$ charges because, starting with such a solution, 
dualities can generate the most general black hole, characterized by $27$ 
conserved charges \cite{Cvetic:1996zq}. The most general black ring in five 
dimensions depends on $27$ charges as well, but also on $27$ dipole charges. 
We will argue that, to generate such general rings, the starting point 
must have $3$ charges and $15$ dipole charges (or {\it vice versa}). Taking
mass and angular momenta into account, the most general black ring would then 
have $21$ parameters. This is a much larger class than those already constructed in the 
BPS case \cite{Elvang:2004ds,Bena:2004de}, and also much larger 
than those previously conjectured in the non-BPS case \cite{Elvang:2004xi}. 

The microscopic interpretation discussed in this paper gives predictions for supergravity solutions that 
have not yet been constructed explicitly:
\begin{enumerate}
\item
The area of a conjectured $9$ parameter family of thermally excited black rings is 
identified.
\item
There exists an $8$ parameter family of {\it extremal} black rings. These black rings 
are not supersymmetric, but they are extremal in the sense that they have vanishing 
temperature.
\item
Thermal black rings are expected to have an inner and an outer horizon, both of 
topology $S^1\times S^2$. The area of the {\it inner} horizon is predicted as well. 
\end{enumerate}

The remainder of this paper is organized as follows. 
In section \ref{sec:duality} we discuss the action of string dualities on black rings
and the parameters needed to describe the most general black ring solution in 
five dimensions. Section \ref{sec:micro} is the core of the paper: we develop the 
microscopic description of the black rings, by adapting the description of black 
strings. In section \ref{sec:thermo} we discuss the resulting
thermodynamics of black rings. We conclude in section \ref{sec:discussion}
with a brief discussion of the ambiguities in the definition of charges.

\section{Black Rings and Dualities}
\label{sec:duality}
At the classical level the theory we discuss is $11$ dimensional supergravity 
compactified on $T^6$ or, equivalently, $N=8$ supergravity in five dimensions. 
The issue we want to address is that, in this theory, duality relates apparently distinct 
configurations, effectively reducing the number of truly 
independent solutions\footnote{We are just discussing
dipole charges, in addition to the conventional black hole charges. 
It is possible that black holes and black rings in five dimensions support 
additional classical hair; indeed, such hair could ultimately account for the 
entire microscopic structure of black holes, as recently advocated by 
Mathur and collaborators \cite{Mathur:2004sv} (and earlier 
in \cite{Larsen:1996ed}).}. In order to factor out this redundancy in the
description we want to determine the duality orbits of black ring solutions.

It is convenient to parametrize solutions in supergravity by their asymptotics near 
infinity. Concretely, this means specifying gravitational mass 
$M$ and angular 
momenta $J_a$ ($a=1,2$), the asymptotic value of all scalar fields 
$X^\infty_I$ ($I=1,\ldots,42$), and the charges of gauge fields. 
The gauge charges $Q_I$ may be computed by gaussian flux integrals\footnote{We 
use units where the 11-dimensional planck length $l_p = (\pi/4G_5)^{1/3}=1$. In these
units all charges are quantized.}
\begin{equation}
Q_I = {1\over 2\pi^2} \int_\infty G_{IJ} {}^*F^I
\end{equation}
In the case of black rings there are also the dipole charges 
$q^I$ specified by integrals
\begin{equation}
q^I = {1\over 2\pi} \int_{S^2} F^I
\end{equation}
where the integral is over a sphere that links the ring. The dipole charges fall off faster 
asymptotically than the ordinary charges, and they are not conserved;
but this does not make them any less
useful in the classification. 

The duality group of $N=8$ supergravity 
in $D=5$ is $E_{6(6)}(R)$. The $42$ scalar fields $X^I$ parametrize the coset 
$E_{6(6)}/USp(8)$ (dimension $78-36=42$). The values of these scalars at
infinity can be chosen arbitrarily; they are just integration constants. Indeed, 
as indicated by the coset form of the scalar manifold, different asymptotic
values of the scalars are related by $E_{6(6)}(R)$ transformations. 
Making a definite choice, {\it e.g.} taking $X^I_\infty$ corresponding to
a square torus with no fluxes and sides of unit length, 
defines the vacuum and breaks the symmetry spontaneously as 
$E_{6(6)}\to USp(8)$.

The $27$ gauge fields in the theory transform in the antisymmetric
symplectic traceless representation of
the $USp(8)$ symmetry leaving the vacuum invariant.  To make
this explicit, it is convenient to organize
the $27$ gauge field charges into the $8\times 8$ central charge 
matrix \cite{Kutasov:1998zh}
\begin{equation}
Z_e = \left( 
\begin{array}{cc}
BJ_{(1)} &  A \\
-A^T & - {1\over 3} BJ_{(3)} + C_{ij}T^{ij} 
 \end{array}
 \right)
 \label{zedef}
\end{equation}
where $J_{(i)}$ are symplectic invariants of $USp(2i)$ and $T^{ij}$ are
a basis of trace-less anti-symmetric $6\times 6$ matrices. One can choose 
the duality frame so that the $2\times 6$ charges $A$ correspond to
$M5$-branes and $KK$-waves wrapped on cycles fully within the $T^6$. 
Then the $B$, $C_{ij}$ are $15$ charges corresponding
to $M2$-branes wrapped on the two-cycles of the $T^6$. 
Acting by the $USp(8)$ duality group, the central charge matrix can be 
skew-diagonalized. In the canonical duality frame just introduced, this 
amounts to turning on just three charges, interpreted as M2-branes 
wrapped on the $(12)$, $(34)$, and $(56)$ cycles of the $T^6$. 

The important point is that a reference solution with just these three charges
is in fact the most general one, up to duality. Let us show explicitly how duality 
can reintroduce all charges: the reference solution, with skew-diagonal
charge matrix, is left invariant by a subgroup $SU(2)^4\subset USp(8)$. New
solutions are thus found by acting with
$USp(8)/SU(2)^4$ on the reference solution . This amounts to adding 
$36-4\times 3= 24$ parameters, 
recovering the general charge configuration with $3+24=27$ 
parameters \cite{Cvetic:1996zq}. 

Next, we discuss the dipole charges, the novel feature introduced
by black rings. These are ``magnetic" charges of the same $27$ gauge fields 
considered above, but they are string-like in character, rather than the point-like 
``electric" charges parametrized in (\ref{zedef}). 
The dipole charges can similarly be organized into a magnetic ``central charge'' matrix
\begin{equation}
Z_m = \left( 
\begin{array}{cc}
B_m J_{(1)} &  A_m \\
-A_m^T & - {1\over 3} B_mJ_{(3)} + C_{m~ij}T^{ij} 
 \end{array}
 \right)
 \label{zmdef}
\end{equation}
which, again, is anti-symmetric and symplectic traceless under
the $USp(8)$ duality in a given vacuum. 
In the canonical duality frame introduced after (\ref{zedef}), 
the $2\times 6$ charges 
$A_m$ are the 6 $M2$-branes and the 6 $KK$-monopoles with one 
direction transverse to the $T^6$; and the $B_m$, $C_{m~ij}$ jointly
describe the 
$M5$-branes wrapping the $15$ independent four-cycles within the $T^6$. 
Again, the magnetic central charge matrix $Z_m$ can be skew-diagonalized by a suitable 
$USp(8)$ duality transformation. However, in the duality frame where the 
electric charge matrix $Z_e$ has already been simplified in this manner, the
magnetic charges are {\it not} in general diagonal. 

The subgroup $SU(2)^4\subset USp(8)$ that leaves the skew-diagonal $Z_e$ 
invariant in general acts non-trivially on $Z_m$, generating 12-parameter 
orbits of equivalent solutions. Concretely, we can choose $A_m=0$
without loss of generality, {\it i.e.} keep only the $15$ dipole $M5$-brane charges.
The $6$ dipole $M2$-branes  and the $6$ dipole $KK$-monopoles can
be taken to vanish because these are generated when $SU(2)^4$ duality transformations 
act on the $M5$-brane dipoles.

In summary, we have shown that the most general black ring in $N=8$ 
supergravity is parametrized up to duality by 21 parameters: the mass $M$, 
2 angular momenta $J_a$ ($a=1,2$), $3$ eigenvalues $Q_i$ ($i=1,2,3$), and 
$15$ dipole charges $B_m$ $C_{m~ij}$. As a check note that the
total number of black ring parameters is $3+27+27=57$ (from gravitational, point-like, 
and string-like charges). Since there are $36$ $USp(8)$ duality parameters
in a given vacuum we find that a seed solution must have $57 - 36 = 21$ parameters, 
as in the more detailed argument. 

In the following we will for simplicity focus on the ``canonical'' $9$ parameter 
family of black rings where the electric and magnetic central charge
matrices
(\ref{zedef})-(\ref{zmdef}) are simultaneously diagonalized. This configuration 
is left invariant by $SU(2)^4\subset USp(8)$ so, when acting on these solutions,
duality transformations can only add the $24$ parameters of the coset 
$USp(8)/SU(2)^4$. The canonical $9$ parameter family of rings therefore correspond, after 
duality is taken into account, to completely general charges, but only three
dipole charges. 

\section{The Microscopic Theory}
\label{sec:micro}
In this section we discuss some features of the microscopic description of the 
black rings. 

\subsection{General Comments}
For extremal black rings it has been proposed that the effective low energy 
theory of the collective excitations is identical to the two dimensional CFT with 
$(4,0)$ supersymmetry governing black holes in {\it four} 
dimensions \cite{Emparan:2004wy,Bena:2004tk,Cyrier:2004hj}. Some motivations for 
this identification are:
\begin{enumerate}
\item
{\it Horizon topology}: the $S^2\times S^1$ topology of the black ring horizon
is identical to that of a five dimensional black string which, in a suitable limit,
is interpreted as a four dimensional black hole. Then the $S^2$ of the ring 
is identified with the horizon of the black hole and the $S^1$ of the ring 
corresponds to the compact dimension.
\item
{\it Near horizon geometry:} in the limit where bulk gravity decouples from the
theory on the branes, the near horizon geometry of the standard black holes is locally
$AdS_3\times S^2$, and the global structure is that of an extremal BTZ black hole. 
AdS/CFT correspondence then identifies the miscroscopic theory 
\cite{Strominger:1997eq}. Black rings similarly allow a near string limit which 
decouples bulk modes and identifies a near horizon BTZ black 
hole \cite{Bena:2004wv, Elvang:2004ds,Bena:2004tk}. This shows that
the two microscopic theories are identical as well.
\item
{\it The effective string}: the usual description of 4D black holes involves an 
asymmetric scaling limit that singles out one compact dimension, which is
interpreted as the spatial direction of an effective string
\cite{Maldacena:1996ds,Larsen:1999dh}. 
It is natural to simply identify this effective string with the black ring. Again, this
is because the geometry near the ring is indistinguishable from that of a straight 
string. 
\end{enumerate}
The working hypothesis of this paper is that 5D black rings can be identified with
5D black strings (and so with 4D black holes) 
also in the non-extremal case, as long as the excitations above 
the extremal limit remains small. This assumption is natural because the arguments 
above for the extremal case remain valid (to the extent we can check them).

\subsection{The CFT description}
In the canonical duality frame described after (\ref{zedef}) the black string 
consists of is  $M5$-branes with wrapping numbers $q^1$, $q^2$, $q^3$
along the cycles orthogonal to the canonical $(12)$, $(34)$, $(56)$ two-cycles.
These five-branes all share one common line which is the locus of the effective 
two dimensional CFT. This spacetime CFT has central charge $c=6q^1 q^2 q^3$ 
for both right and left movers and $(4,0)$ supersymmetry. The $M2$ branes 
(and all other charges) are realized as charged excitations of this theory. 

The supersymmetric sector of the CFT are the right movers. The $N=4$ superconformal 
algebra contain an affine $SU(2)$ current at level $k = {\hat c}= {2\over 3}c= 4q^1 q^2 q^3$. 
States with quantum number $J$ under a $U(1)$ subgroup of this R-symmetry have level 
\begin{equation}
h^{\rm rot}_R = {1\over k}j^2 = {1\over 4q^1 q^2 q^3}J^2
\end{equation}
The $SU(2)$ symmetry is interpreted in spacetime as the rotation group in the three dimensional
transverse space; so the quantum number $J$ is simply the projection of the angular momentum
along the quantization axis.

There is one additional right moving current. This current carries the charges dual
to the charges defining the background. Concretely, spacetime supersymmetry
gives the BPS mass\footnote{In our units the brane tensions are automatically taken
into account correctly.}
\begin{equation}
M^2 = (q^1 X_1 + q^2 X_2 + q^3 X_3)^2 R^2 + (Q_1 X^1  + Q_2 X^2  + Q_3 X^3)^2
\label{BPSmass}
\end{equation}
where $X_I$ parametrize the volumes of the four-cycles wrapped by the $M5$-branes,
$R$ is the radius of the direction along the effective string, and
the $X^I=1/X_I$ correspond to the volumes of the dual two-cycles wrapped by the 
$M2$-branes.
Treating the $M5$-branes as a heavy background, the energy associated with the 
$M2$-brane excitations is
\begin{equation}
\Delta M \simeq 
{1\over 2(q^1X_1 + q^2 X_2 + q^3 X_3)R} (Q_1X^1 + Q_2 X^2 + Q_3 X^3)^2 
\end{equation}
In the decoupling limit where the CFT applies this equality becomes exact.
If excitations with this energy arise as zero-modes of affine currents the
conformal weight associated with the charge is
\begin{eqnarray}
h^{\rm M2}_R &=& {R\over 2}\Delta M  = 
{1\over 12q^1 q^2 q^3} (Q_1q^1  + Q_2 q^2  + Q_3 q^3)^2 
\label{hRM2}
\end{eqnarray}
where we let the scalar fields attain
their attractor values $X^I= q^I/(q^1 q^2 q^3)^{1/3}$ for $I=1,2,3$ (for a discussion of 5D 
attractors emphasizing black rings see \cite{Kraus:2005gh}). 

The remaining two linear combinations of $M2$-brane charges are not affected 
by spacetime supersymmetry; so these are carried by left moving currents. 
The levels of these currents are constrained by modular invariance of the CFT
which relates right and left moving currents. 
The simplest possibility is to form a lattice of signature $(2,2)$ with the right 
moving currents \cite{Larsen:1999dh}. This prescription gives the conformal weights of 
the zero-modes
\begin{eqnarray}
h^{\rm M2}_L &=& {1\over 4q^1 q^2 q^3} (Q_1q^1 - Q_2 q^2)^2 + 
{1\over 12q^1 q^2 q^3} (Q_1q^1  + Q_2 q^2  - 2 Q_3 q^3)^2\nonumber\\
&=& {1\over 3q^1 q^2 q^3} [ (Q_1q^1)^2  +  (Q_2q^2)^2  + 
 (Q_3q^3)^2  ] \nonumber\\
 &&-  {1\over 3q^1 q^2 q^3} [ Q_1q^1 \cdot Q_2q^2 +  Q_1q^1\cdot Q_3q^3+ 
 Q_2q^2 \cdot Q_3q^3]
  \label{hLM2}
\end{eqnarray}
Since (\ref{hLM2}) is not protected by supersymmetry this expression can only be trusted in
the semi-classical regime.

It is not difficult to generalize the argument leading to (\ref{hRM2}) and (\ref{hLM2}) to 
find a lattice of signature $(14,14)$ that describes all $27$ charged excitations 
and angular momentum. The result requires a bit more notation because
it depends on the complex structure moduli so we just refer to \cite{Larsen:1999dh}
for the details. The microscopic theory describes the most general thermal
black ring up to duality when this full set of excitations is taken into account.

The arguments presented for the existence of the various currents and for the weight 
of their zero-modes is somewhat heuristic; but there are several checks on the 
assignments:
\begin{enumerate}
\item
{\it Triality}: the formulae (\ref{hRM2}) and (\ref{hLM2}) for the weights are invariant
under simultaneous permutations of the $M5$ brane charges and the $M2$ brane charges. 
This symmetry is in the Weyl subgroup of the duality group and must be respected.
\item
{\it Rational levels}: the levels (\ref{hRM2}) and  (\ref{hLM2}) turned out to be rational
even though the attractor mechanism fixes the scalars at irrational values
$X^I= q^I/(q^1 q^2 q^3)^{1/3}$. 
\item
{\it Level matching}: the difference between levels
\begin{eqnarray}
h^{\rm M2}_L - h^{\rm M2}_R  &= &{1\over 4q^1 q^2 q^3} [ (Q_1q^1)^2  +  (Q_2q^2)^2  + 
 (Q_3q^3)^2  ]\\
 &&-  {1\over 2q^1 q^2 q^3} [ Q_1q^1 \cdot Q_2q^2 +  Q_1q^1 \cdot Q_3q^3+
  Q_2q^2 \cdot Q_3q^3] \nonumber
\end{eqnarray}
is quantized in the same unit $1/k=1/4q^1 q^2 q^3$ as the angular momentum. This
means the higher modes of currents can be matched consistently. 
\item
{\it Global symmetry}: the total set of $27$ currents in the theory (discussed after (\ref{hLM2})
and in section 2) transform as 
$({\bf 1},{\bf 1})\otimes ({\bf 2},{\bf 6})\otimes ({\bf 1},{\bf 14})$ under the duality 
$SU(2)\times USp(6)\subset USp(8)$ preserved by the fixed scalar conditions. 
The $({\bf 1},{\bf 1})\otimes ({\bf 2},{\bf 6})$ are right movers, with level 
determined by the spacetime BPS algebra, as discussed before (\ref{hRM2}). 
The $({\bf 1},{\bf 14})$ relates the normalizations of {\it all} left moving 
currents \cite{Larsen:1999dh} and confirm (\ref{hLM2}). 
\end{enumerate}

\subsection{Results}
We now count the entropy from the degeneracy of the states in the CFT. The vertex 
operators of the states we are counting are given by
$$
{\cal V}_{\rm tot} = {\cal V}_{\rm irr} {\cal V}_{U(1)}
$$
The ${\cal V}_{U(1)}$ is constructed from the $U(1)$ currents such that the full
vertex operator carries the correct $U(1)$ charge. The conformal weight accounted
for by this was derived above. The ${\cal V}_{\rm irr}$ can be specified freely and so 
gives rise to entropy. The expression for the entropy is given as usual by Cardy's formula
\begin{equation}
S = 2\pi \left[ \sqrt{ch^{\rm irr}_L \over 6} + \sqrt{ch^{\rm irr}_R \over 6}\right]
\label{micent}
\end{equation}
where presently the irreducible weights are given
\begin{eqnarray}
h_L^{\rm irr} &=& {\epsilon+p\over 2}
-{1\over 3q^1 q^2 q^3}\left[ (Q_1 q^1)^2  + (Q_2 q^2)^2 + (Q_3 q^3)^2 \right]  \nonumber \\
&+&{1\over 3q^1 q^2 q^3}\left[ Q_1 q^1\cdot Q_2 q^2 + Q_1 q^1 \cdot Q_3 q^3
+ Q_2 q^2 \cdot Q_3 q^3   \right] \label{hL} \\
h_R^{\rm irr} &=& {\epsilon-p\over 2} -{1\over 12q^1 q^2 q^3}\left[ Q_1 q^1+ Q_2 q^2 + Q_3 q^3\right]^2  
-{1\over 4q^1 q^2 q^3} J^2  \label{hR} 
\end{eqnarray}
This is our final result for the entropy. In this formula the momentum quantum number
$p$ along the
black string should be identified with the angular momentum along the black ring
and the four-dimensional angular momentum $J$ is the angular 
momentum transverse to the ring $J=J_\phi$. 

The extremal limit is given by $h_R^{\rm irr}=0$. In this limit we can eliminate $\epsilon$ 
and find
\begin{eqnarray}
h_L^{\rm irr} &=& p
-{1\over 4q^1 q^2 q^3}\left[ (Q_1 q^1)^2  + (Q_2 q^2)^2 + (Q_3 q^3)^2 \right]  \\
&+&{1\over 2q^1 q^2 q^3}\left[ Q_1 q^1\cdot Q_2 q^2 + Q_1 q^1\cdot Q_3 q^3
+ Q_2 q^2 \cdot Q_3 q^3  \right] + {1\over 4q^1 q^2 q^3} J^2  \nonumber
\label{hirrext}
\end{eqnarray}
If in addition we impose supersymmetry then $J=0$. In this BPS limit 
our general result (\ref{micent}) for the entropy reduces to the one deduced 
from $E_{7(7)}$ duality symmetry of the black 
string in \cite{Bena:2004tk,Kallosh:1996uy}, and from counting deformations of $M5$-branes 
in \cite{Cyrier:2004hj,Maldacena:1997de}. The 8 parameter family of configurations with 
$h^{\rm irr}_R=0$ but $J\neq 0$ corresponds to black rings that are extremal, 
but not supersymmetric. 

The general 9-parameter family of near extremal black rings have not yet been
constructed in supergravity. The formula (\ref{micent}) for their entropy predicts
the area of their outer horizon
\begin{equation}
A_+ = 2\pi^2 ~ {1\over 2}(\sqrt{h^{\rm irr}_L} + \sqrt{h^{\rm irr}_R})
\end{equation}
These black rings are expected to have an inner 
horizon as well, also of topology $S^1\times S^2$ \cite{Elvang:2003yy,Elvang:2004xi}. 
The considerations of \cite{Cvetic:1997uw} predicts the area of the {\it inner} horizon 
\begin{equation}
A_- = 2\pi^2 ~ {1\over 2}(\sqrt{h^{\rm irr}_L} - \sqrt{h^{\rm irr}_R})
\end{equation}
where the $h_{L,R}^{\rm irr}$ are given in (\ref{hL}) and (\ref{hR}).

\section{Black Ring Thermodynamics}
\label{sec:thermo}
In this section we derive the thermodynamics of the near extremal black ring. 
The strategy is to treat the entropy (\ref{micent}) as a potential that 
generates all other physical features of the ring through the first law of thermodynamics 
\begin{equation}
dM = TdS + \Omega_a dJ^a + \Phi^I dQ_I + \varphi_I dq^I
\end{equation}
Here $M$ is the total mass of the ring, $T$ is the temperature of Hawking radiation
and $S$ is the entropy. The $J^a$ with $a=1,2$ are the two angular 
momenta and the $\Omega_a$ are the corresponding potentials, interpreted geometrically
as the rotational velocities at the horizon. The M2-brane charges are $Q_I$, and the $\Phi^I$
are the corresponding electromagnetic potentials at the horizon. Finally the M5-brane
dipole charges are $q^I$, and $\varphi_I$ are the corresponding magnetic potentials
at the horizon.

The first law readily gives the temperature through
\begin{equation}
{1\over T} = \left( {\partial S\over \partial M}\right)_{J_a,Q_I,q^I}
\end{equation}
We find
\begin{equation}
{1\over T} = {1\over T_R} + {1\over T_L}
\end{equation}
where
\begin{equation}
T_{L,R} = {2\over \pi} \sqrt{6h^{\rm irr}_{L,R}\over c}
\end{equation}
The two temperatures $T_L$ and $T_R$ are interpreted as usual as the independent 
temperatures of the left and right moving excitations. 

In the extremal limit $h^{\rm irr}_R\to 0$ we have $T_R\to 0$ so 
that right moving excitations are forced to the ground state. This takes the spacetime 
temperature $T\to 0$ as well. On the other hand, the left moving temperature 
approaches the finite value
\begin{equation}
T_L \to {12\over \pi c} \sqrt{ch^{\rm irr}_{L}\over 6} = {6\over\pi^2c}S
\end{equation}
in the extremal limit. This is the temperature of the highly degenerate ground state 
responsible for the entropy.

A natural parametrization for the two angular momenta $a=1,2$ is to identify one
index with the momentum along the ring $a=p$ and the other with the angular 
momentum $J_\phi=J$ in the plane transverse to the ring. We then find
\begin{equation}
{\Omega_p\over T} = -\left( {\partial S\over \partial p}\right)_{M,J_\phi,Q_I,q^I} = {\pi\over 2} 
\sqrt{c\over 6}\left[ {1\over\sqrt{h^{\rm irr}_R}} - {1\over\sqrt{h^{\rm irr}_L}}\right]
= {1\over T_R}- {1\over T_L}
\end{equation}
for the rotational velocity along the ring; and
\begin{equation}
{\Omega_\phi\over T} = -\left( {\partial S\over \partial J}\right)_{M,p,Q_I,q^I}
= {\pi\over 2} \sqrt{c\over 6h^{\rm irr}_L} {J\over q^1 q^2 q^3} = {\pi\over 2} \sqrt{6\over ch^{\rm irr}_L} J
= \pi^2 ~{J\over S_L}
\end{equation}
for the rotational velocity transverse to the ring. Here $S_L$ is the entropy of the left-movers.

The rotational velocity can never exceed the speed of light $\Omega_p<1$ as this would,
effectively, amount to the development of closed time-like curves. However, in the extremal limit, 
the rotational velocity along the direction of the ring approaches the speed of light as
\begin{equation}
\Omega_p = {T_L-T_R\over T_L +T_R}\to 1^-
\end{equation}
On the other hand, the rotational velocity in the plane transverse to the ring
slows down $\Omega_\phi\to 0$. This happens at the same rate as $T\to 0$ 
such that, in the extremal limit, there can be a finite ratio
\begin{equation}
{\Omega_\phi\over T}\to  \pi^2 ~{J\over S}
\end{equation}
Of course BPS black holes have $J=0$.

For completeness, let us also consider the electromagnetic potentials. They are
\begin{eqnarray}
{\Phi^1\over T} &=& -\left( {\partial S\over \partial Q_1}\right)_{M,J_a,Q_{2,3},q^I} \\ &=& 
{\pi\over 3q^1 q^2 q^3} \sqrt{c\over 6} \left( {2Q_1 q^1 - Q_2 q^2 - Q_3 q^3\over \sqrt{h^{\rm irr}_L}}+
{ Q_1 q^1 + Q_2 q^2 + Q_3 q^3 \over 2\sqrt{h^{\rm irr}_R}}\right) q^1 \nonumber\\
&=& {4\over c} \left( {2Q_1 q^1 - Q_2 q^2 - Q_3 q^3\over T_L}+
{ Q_1 q^1 + Q_2 q^2 + Q_3 q^3 \over 2T_R}\right) q^1
 \nonumber
\end{eqnarray}
for the usual charges and
\begin{eqnarray}
{\varphi_1\over T} &=& -\left( {\partial S\over \partial q^1}\right)_{M,J_a,Q_I,q^{2,3}} \\ &=& 
{\pi\over 3q^1 q^2 q^3} \sqrt{c\over 6} \left( {2Q_1 q^1 - Q_2 q^2 - Q_3 q^3\over \sqrt{h^{\rm irr}_L}}+
{ Q_1 q^1 + Q_2 q^2 + Q_3 q^3 \over 2\sqrt{h^{\rm irr}_R}}\right) Q^1 \nonumber\\
&~&~~~~~~+ \pi  \left( \sqrt{c\over 6h^{\rm irr}_L}{\epsilon+p\over 2q^1}  + 
\sqrt{c\over 6h^{\rm irr}_R}{\epsilon-p\over 2q^1} \right)\nonumber\\
&=& {4\over c} \left( {2Q_1 q^1 - Q_2 q^2 - Q_3 q^3\over T_L}+
{ Q_1 q^1 + Q_2 q^2 + Q_3 q^3 \over 2T_R}\right) Q_1 + {\epsilon-\Omega_p p\over q^1T}
 \nonumber
\end{eqnarray}
for the dipole charges. The remaining potentials $\Phi^{2,3}$ and $\varphi_{2,3}$
are found by the obvious cyclic permutations. 

In the extremal limit $T_R\to 0$ and $\epsilon\to p$ with $T_L$ finite so that
\begin{eqnarray}
\Phi^I &\to & {4\over c} (Q_J q^J)~ q^I \\
\varphi_I& \to & {4\over c} (Q_J q^J)~ Q_I
\end{eqnarray}
Note that, in the extremal limit, the electric and magnetic potentials are related such that 
\begin{equation}
\Phi^I Q_I= \varphi_I q^I~~~~~({\rm no~sum~over~I})
\end{equation}

\section{Discussion} 
\label{sec:discussion}
As we have emphasized repeatedly, the strategy in this paper is to find a microscopic
theory of black rings by identifying the circular black rings with straight black strings. The results
are formulae for the thermodynamics of objects that seem quite difficult to construct 
even at the level of supergravity. As is clear from the derivation, the results are 
written in the variables that are natural for the black string. When the results
are interpreted in terms of the black ring, as advocated here, these variables must be 
defined in the near horizon geometry; and so they may in general differ from the 
corresponding variables in the asymptotic geometry.

The distinction between near horizon and asymptotic variables is familiar already from 
the extremal case. The near horizon charges $Q_I$ used in the present paper can be 
identified with the ``barred" charges ${\bar Q}_I$ of \cite{Bena:2004tk,Bena:2005ni} and 
should be distinguished from asympotic charges of the 
asymptotically flat black ring \cite{Elvang:2004ds,Cyrier:2004hj}. 
Similarly, as in \cite{Bena:2004tk,Bena:2005ni}, the momentum along the ring 
is related to the angular momenta in the asymptotic space as $p = -(J_\psi+J_\phi)$ 
(in the notation of \cite{Elvang:2004ds}). 

For the thermal ring there is an additional ambiguity: the excitation energy of the 
microscopic theory plays the role of  ``mass" in the near horizon geometry but this 
mass, and its dual temperature, cannot be identified with the mass parameter in 
the asymptotically flat space. The distinction is seen clearly by noting
that the black ring description associates a non-vanishing energy with {\it all} 
charged excitations, {\it e.g.} in (\ref{BPSmass}). In contrast, the BPS mass in 
the asymptotically flat space is independent of dipole charges, through the
mechanism familiar from supertubes \cite{Mateos:2001qs}.

The microscopic definitions of the parameters used here are clear close to 
the horizon: the near horizon geometry of the ring (not yet constructed explicitly) is 
expected to be $AdS_3$, because this is the geometry of the corresponding string.
The charges, angular momenta, and mass are the parameters defined asymptotically 
in this space; and this is also where the central charge of the dual theory is defined, 
as are the levels of the various affine currents exploited in this paper. 

It would clearly be interesting to complement the description pursued here
with an interpretation of black rings as excitations in the asymptotically
flat geometry. This would also help clarify the relation between black rings and
black holes in five dimensions.

\section*{Acknowledgements}
I thank I. Bena, J. de Boer, H. Elvang, P. Kraus, D. Mateos, H. Reall, and E. Verlinde for discussions.
This work was supported in part by the US DoE under grant
DE-FG02-95ER40899.


\end{document}